\documentclass[aps,twocolumn,floatfix]{revtex4}
\usepackage{graphicx} 
\usepackage{amsmath}
\usepackage{color}
\newcommand{\bea}{\begin{eqnarray}}
\newcommand{\eea}{\end{eqnarray}}
\newcommand{\bse}{\begin{subequations}}
\newcommand{\ese}{\end{subequations}}

\begin{document}

\title{Noninteracting Electrons in a Prototypical One-Dimensional Sinusoidal Potential}

\author{David C. Johnston}
\affiliation{Ames Laboratory and Department of Physics and Astronomy, Iowa State University, Ames, Iowa 50011, USA}

\date{\today}

\begin{abstract}

A prototypical model of a one-dimensional metallic monatomic solid containing noninteracting electrons is studied, where the argument of the cosine potential energy periodic with the lattice contains the first reciprocal lattice vector $G_1=2\pi/a$, where $a$ is the lattice constant.  The time-independent Schr\"odinger equation can be written in reduced variables as a Mathieu equation for which numerically-exact solutions for the band structure and wave functions are obtained.  The band structure has band gaps that increase with increasing amplitude~$q$ of the cosine potential.  In the extended-zone scheme, the energy gaps decrease with increasing index $n$ of the Brillouin-zone boundary $ka = n\pi$ where $k$ is the crystal momentum of the electron.  The wave functions of the band electron are derived for various combinations of $k$ and $q$ as complex combinations of the real Mathieu functions with even and odd parity and the normalization factor is discussed.  The wave functions at the bottoms and tops of the bands are found to be real or imaginary, respectively, corresponding to standing waves at these energies.  Irrespective of the wave vector~$k$ within the first Brillouin zone, the electron probability density is found to be periodic with the lattice.  The Fourier components of the wave functions are derived versus~$q$, which reveal multiple reciprocal-lattice-vector components with variable amplitudes in the wave functions unless $q=0$.  The magnitudes of the Fourier components are found to decrease exponentially as a power of $n$ for $n\sim 3$ to 45 for $ka=\pi/2$ and $q=2$  and a precise fit is obtained to the data.  The probability densities and probability currents obtained from the wave functions are also discussed. The probability currents are found to be zero for crystal momenta at the tops and bottoms of the energy bands, because the wave functions for these crystal momenta are standing waves.   Finally, the band structure is calculated from the central equation and compared to the numerically-exact band structure.

\end{abstract}

\maketitle

\section{Introduction}

Many features of the properties of two-and three-dimensional metallic crystals appear in the study of one-dimensional solids containing noninteracting electrons.  The earliest such model is the Kronig-Penney model with a periodic square-well potential, a limiting form of which is the periodic Dirac-comb potential~\cite{Kronig1931}.  However, more realistic potential energies are obtained from superpositions of sinusoidal terms containing arguments with different reciprocal-lattice vectors, where the resultant potential is periodic with the lattice by construction.  In this paper we consider the simplest case where the potential energy is proportional to $\cos(G_1x)$, where $G_n=n2\pi/a$ with $n=1$ which is the lowest-order reciprocal-lattice vector where $a$ is the lattice constant~\cite{Slater1952}. After defining reduced variables, the resultant time-independent Schr\"odinger equation is the so-called Mathieu equation~\cite{McLachlan1951, Pipes1953, Kokkorakis2000, Choun2015} for which numerically-exact solutions for the dispersion relations (band structure) and wave functions versus crystal momentum~$k$ can be obtained using recent additions to the {\tt Mathematica} program suite of special functions~\cite{Mathematica}.  Previously, the band structures and wave functions and associated quantities were only briefly illustrated~\cite{Carver1971}.  The applications of the Mathieu equation to other problems have also been considered~\cite{Ruby1996, Horne1999, GutierrezVega2003}.

Here we utilize {\tt Mathematica} to calculate to high accuracy the band structures, wave functions, and probability densities versus position and amplitude of the cosine potential.  Some of our wave functions are similar to those in Ref.~\cite{Horne1999}.  In addition, we obtained Fourier series spectra of the $G_n$ components in the wave functions up to $n=45$ for a particular value of the cosine potential amplitude.  These high-frequency components are present even when only the $n=1$ reciprocal-lattice vector is present in the potential energy because pure sinusoidal wave functions are not solutions to the Mathieu Schr\"odinger equation except for $q=0$ and contain higher-order contributions that generally decrease in amplitude with increasing order~$n$.  The probability amplitudes and currents are calculated for representative crystal momenta in the first, second, and third Brillouin zones.  The probability currents are found to be zero for wave functions at the tops and bottoms of the energy bands because these states are standing waves.

The background needed for the calculations and the associated notation are given in Sec.~\ref{Sec:Basics}.  In Sec.~\ref{Sec:BandSTructGaps} the band structures for several values of the amplitude of the cosine potential energy are calculated.  Then the corresponding wave functions versus position and their Fourier components are presented in Sec.~\ref{Sec:WaveFcns}.   The probability densities and probability currents are discussed in Sec.~\ref{PJ}.  The band structure is calculated from the central equation in Sec.~\ref{Sec:CentEq} and compared with the numerically-exact results.  Concluding remarks are given in Sec.~\ref{Sec:Conclusion}.

\section{\label{Sec:Basics} Background}

The time-independent Schr\"odinger equation in one dimension ($\equiv x$) for the wave function $\psi(x)$, potential energy $U(x)$ and energy $E$ of a particle is
\bea
-\frac{\hbar^2}{2m}\frac{d^2\psi(x)}{dx^2}+U(x)\psi(x) = E\psi(x).
\label{Eq:Sch222}
\eea
A real potential energy that is periodic with the lattice satisfies
\bea
U(x) = \sum_{n=1}^\infty \left[U_{n{\rm c}} \cos\left(\frac{n2\pi}{a}x\right) + U_{n{\rm s}} \sin\left(\frac{n2\pi}{a}x\right)\right]\label{Eq:U(x)1},\nonumber\\
\eea
where $n$ is a positive integer and $U_{n{\rm c}}$ and $U_{n{\rm s}}$ are real coefficients.  In one dimension the values of $n 2\pi/a$ are just the magnitudes of the reciprocal-lattice vectors \mbox{${\bf G}_n = [n 2\pi/a]\hat{\bf i}$}.  Thus $U(x)$ can be expressed as  a Fourier series in the reciprocal lattice vectors.  If all $U_n$ are zero, one has the Schr\"odinger equation for a free electron with wave function~$\psi(x) $ and energy~$E$ given by
\bse
\label{Eqs:psixABkxx}
\bea
\psi(x) &=& Ae^{ik_xx}+Be^{-ik_xx},\label{Eq:psiFE}\\
E &=& \frac{\hbar^2k^2}{2m},\label{Eq:EFE}
\eea
\ese
where $A$ and $B$ are arbitrary real coefficients.

As noted above, here we consider a sinusoidal potential energy $U(x)$ with amplitude $U_1$ containing only the first reciprocal lattice vector ${\bf G}_1=(2\pi/a)\,\hat{\bf i}$ given by
\bea
U(x) = U_1\cos\left(2\pi\frac{x}{a}\right).
\label{Eq:U(x)}
\eea
Then the Schr\"odinger equation~(\ref{Eq:Sch222}) becomes
\bea
-\frac{\hbar^2}{2m}\frac{d^2\psi(x)}{dx^2} + U_1\cos(2\pi x/a)\psi(x) = E\psi(x).
\label{Eq:Sch33}
\eea
Defining
\bea
x_a=x/a,
\label{Eq:xaDef123}
\eea
we have
\bea
U(x_a) = U_1\cos(2\pi x_a).
\eea
Using Eq.~(\ref{Eq:xaDef123}), Eq.~(\ref{Eq:Sch33}) can be written
\bea
\frac{d^2\psi(x_a)}{dx_a^2} + \left[\frac{2mEa^2}{\hbar^2} - \frac{2mU_1a^2}{\hbar^2}\cos(2\pi x_a)\right] \psi(x_a) = 0.\nonumber\\
\label{Eq:Sch56}
\eea
If the potential energy coefficient $U_1=0$, one obtains the free-electron results~(\ref{Eqs:psixABkxx}).

Defining the dimensionless reduced parameters
\bea
u_1\equiv \frac{2mU_1a^2}{\hbar^2},\quad \varepsilon \equiv  \frac{2mEa^2}{\hbar^2},\quad z = \pi x_a,
\label{Eq:u1Def}
\eea
Eq.~(\ref{Eq:Sch56}) becomes
\bea
\frac{d^2\psi(z)}{dz^2} +\left[\frac{\varepsilon}{\pi^2}-\frac{u_1}{\pi^2}\cos(2z)\right] \psi(z) =0.
\label{Eq:Sch59}
\eea
For free electrons ($u_1=0$), the wave function and eigenenergy are given by Eqs.~(\ref{Eqs:psixABkxx}), for which the dispersion relation in reduced variables is
\bea
\frac{\varepsilon}{\pi^2} = \left(\frac{ka}{\pi}\right)^2\quad{\rm or} \quad\varepsilon=(ka)^2.
\label{Eq:DispRelnFERedVar}
\eea
Finally, defining the variables
\bea
a_1 = \frac{\varepsilon}{\pi^2},\qquad q = \frac{u_1}{2\pi^2},
\label{Eqs:a1q}
\eea
Eq.~(\ref{Eq:Sch59}) reads
\bea
\frac{d^2\psi(z)}{dz^2} +\left[a_1- 2q\cos(2z)\right] \psi(z) =0.
\label{Eq:Sch99}
\eea

This differential equation is the \mbox{Mathieu} equation for which numerically-exact solutions for even- and odd-parity wave functions $\psi(z)$  called \mbox{Mathieu} functions can be obtained versus $a_1$ and~$q$ using {\tt Mathematica}~\cite{Mathematica}.  For $q=0$ the \mbox{Mathieu} functions are $\psi(x) = C\cos(kx)$ and $D\sin(kx)$, respectively, which can be combined to form the free-electron wave function in Eq.~(\ref{Eq:psiFE}).

A special case of the solution of Eq.~(\ref{Eq:Sch99}) for the wave function with real energy $a_1$ and~cosine amplitude $2q$, where the wave function satisfies the Bloch theorem applied to the case of our potential energy periodic in the lattice parameter~$a$,  is~\cite{Mathematica, Cottey1971}
\bea
\psi(x_a) = e^{i\,r\,\pi x_a}f(x_a),
\label{Eq:psi(z)}
\eea
where
\bea
r = \frac{ka}{\pi}
\eea
and $f(x_a)$ is a function with periodicity~1. Hence this periodicity is the same as that of the lattice and of $U(x_a)$ in Eq.~(\ref{Eq:U(x)}).  In Eq.~(\ref{Eq:Sch99}), the value of
\bea
a_1 = \frac{\varepsilon}{\pi^2} \equiv a_r(q)
\label{Eq:a1Def}
\eea
depends on the values of $r$ and $q$, where $a_r(q)$ is the ``characteristic value'' of~$a_1$ for given values of $r$ and~$q$.  

\section{\label{Sec:BandSTructGaps} Band Structures and Band Gaps}

\begin{figure}
\includegraphics[width=3.in]{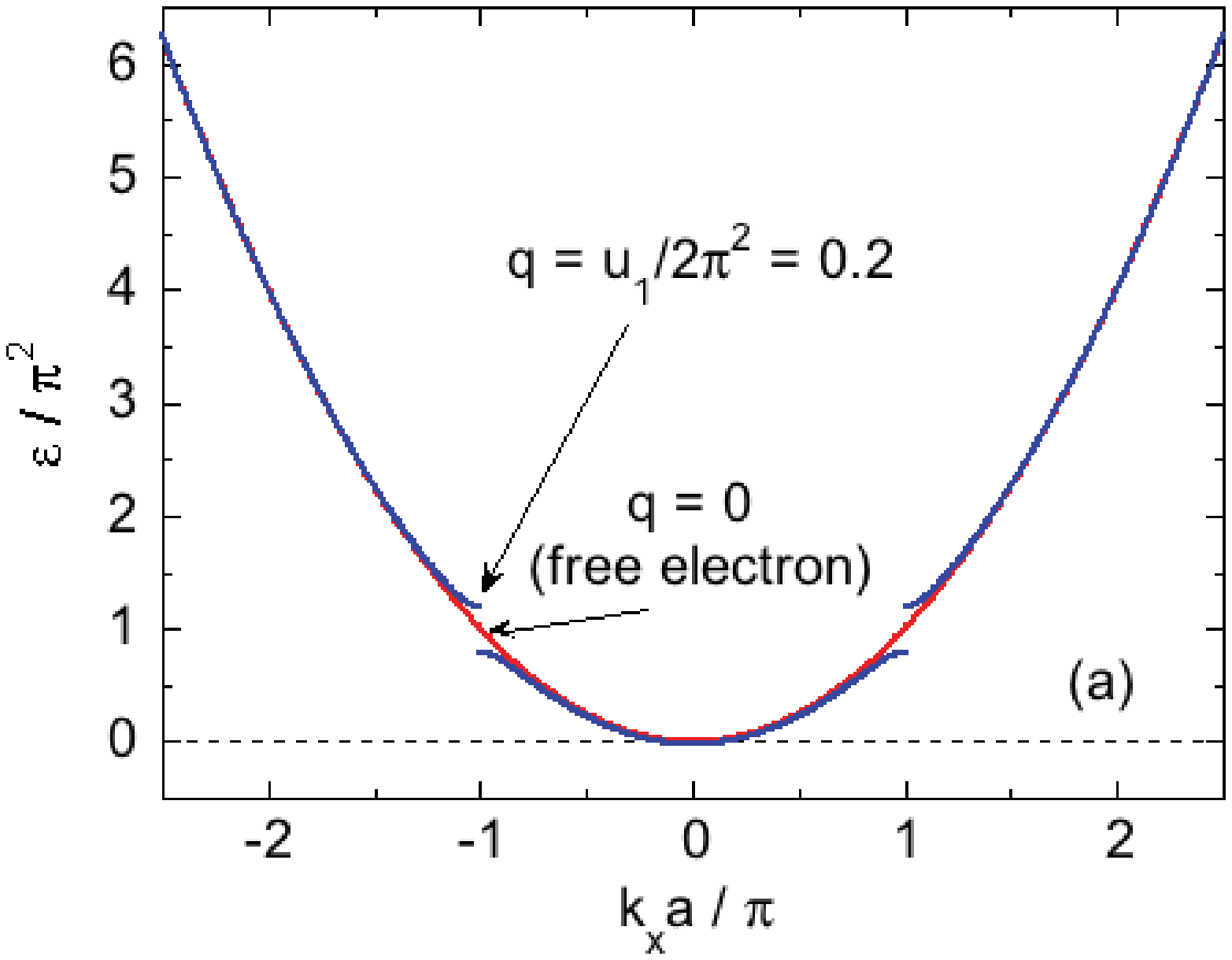}
\includegraphics[width=3.in]{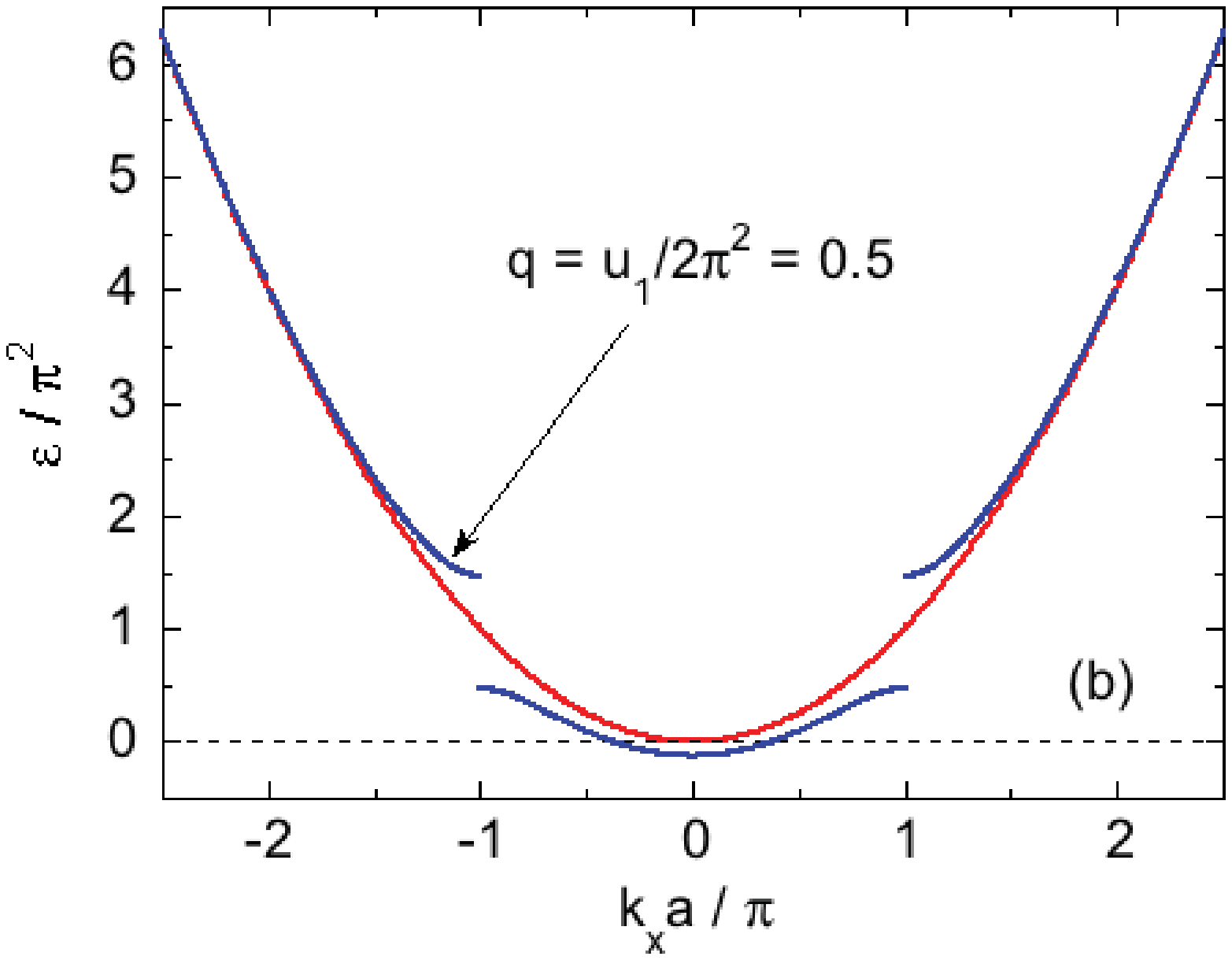}
\includegraphics[width=3.in]{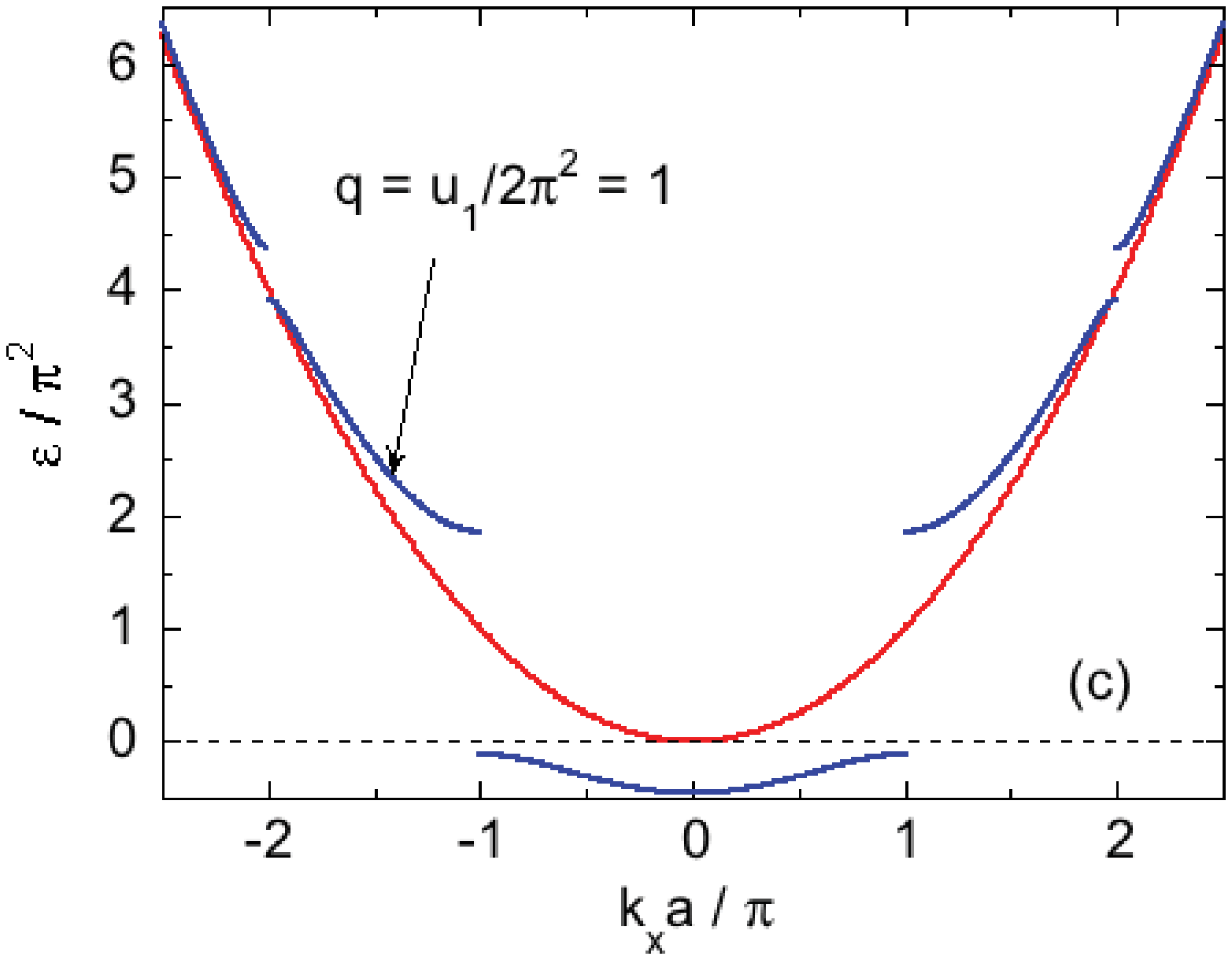}
\caption{Band Structure $\varepsilon/\pi^2$ versus $k_xa/\pi$ in the extended-zone scheme for the first three bands (blue)  associated with the sinusoidal potential~(\ref{Eq:U(x)}) with the wave vector $2\pi/a$. Vertical band gaps occur for the bands at Brillouin-zone boundaries $k_xa = \pm n\pi$ with integer~$n=1$ and~2 where the amplitude $q$ of the potential in Eq.~(\ref{Eq:Sch99}) is (a)~0.2, (b)~0.5, and~(c)~1.  Band gaps at $k_xa/\pi =  \pm2$ are present for $q=0.2$ and~0.5 but are too small to see clearly in panels~(a) and~(b), respectively.  Also shown in each panel for comparison is the dispersion relation $\varepsilon/\pi^2 = (k_xa/\pi)^2$ for free electrons from Eq.~(\ref{Eq:DispRelnFERedVar}) (red curves).  The free-electron band passes through the band gaps at approximately the middle of the two band gaps shown.}
\label{Fig:Matthieu_band_struct}
\end{figure}

All calculations in this paper were carried out using {\tt Mathematica} in which the Mathieu and related functions are built in~\cite{Mathematica}.    Figure~\ref{Fig:Matthieu_band_struct} shows the band structure for amplitudes  $q = 0.2,$ 0.5, and~1 of the cosine potential energy in Eq.~(\ref{Eq:Sch99}), where energy gaps appear in the band structure.  The free-electron dispersion~(\ref{Eq:DispRelnFERedVar}) is shown by the red curves.  Furthermore, the lowest-energy band is negative over part or all of the band.  As $q$ increases from zero, first only the lower part of the band is negative, but eventually at $q=1$ all states in the band have negative energies.

In addition to the expected band gap at $ka/\pi=1$, band gaps also occur at higher-order Brillouin zone boundaries. These arise because the wave function solution of the Schr\"odinger equation for a band-electron wave vector at $n=1= ka/\pi = G_1a/2\pi$ contains additional reciprocal-lattice vector components (see following Sec.~\ref{Sec:WaveFcns}) except for the free-electron case with $q=0$.

\begin{figure}
\includegraphics[width=3.in]{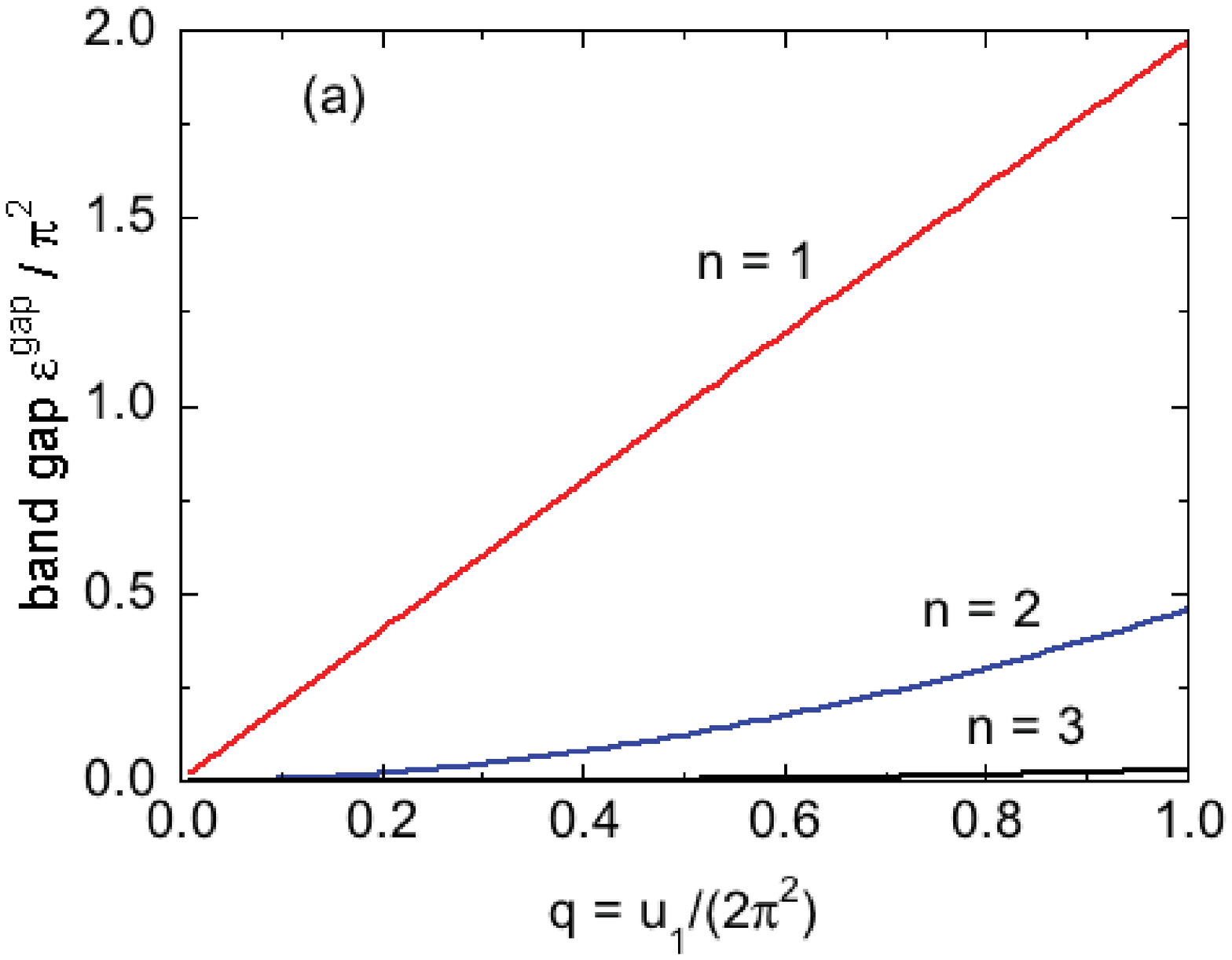}
\includegraphics[width=3.in]{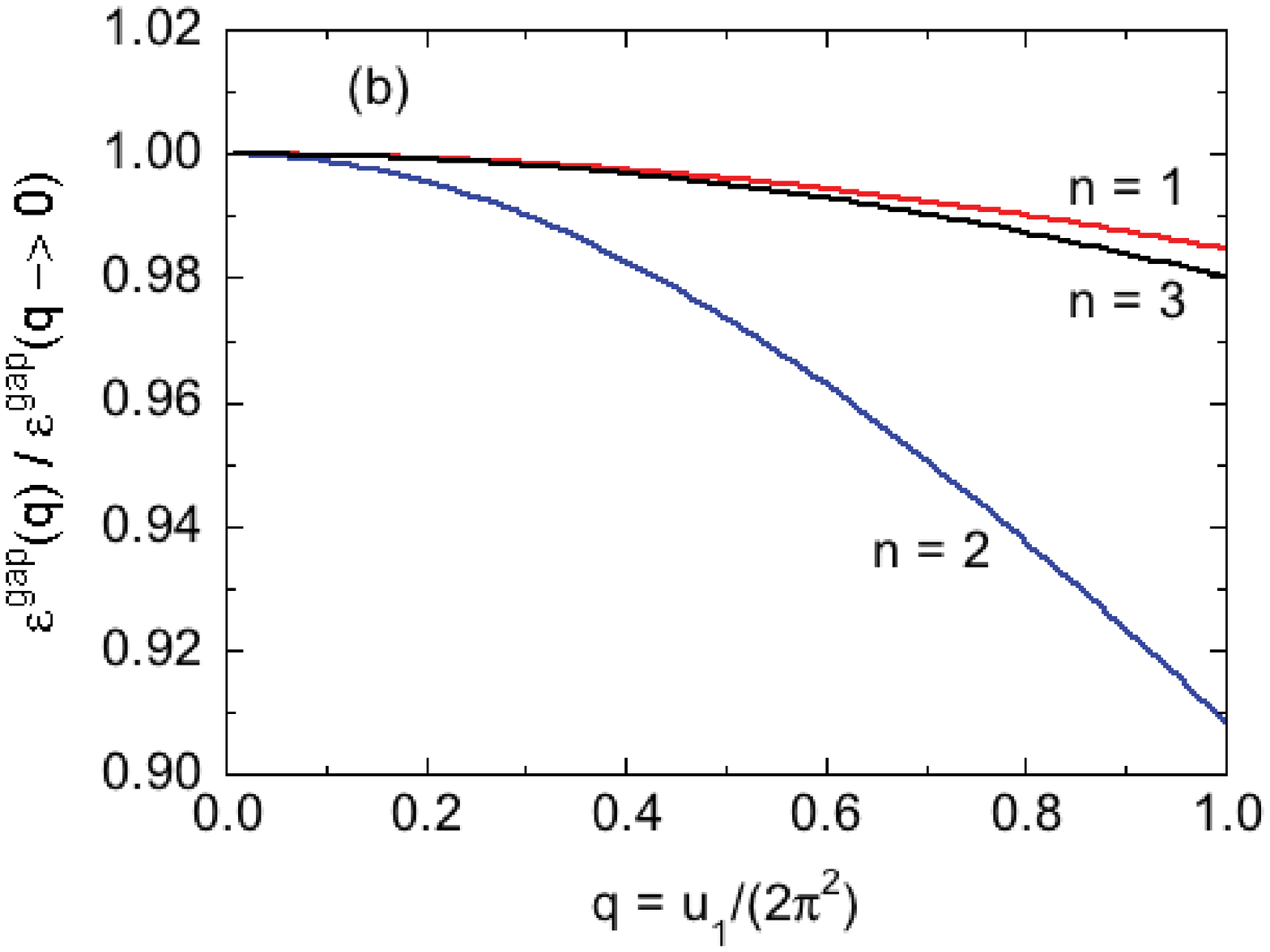}
\caption{(a)~Reduced band gaps $\varepsilon^{\rm gap}/\pi^2$ at the first, second, and third Brillouin zone boundary indices $n=G_na/\pi=1$, 2, and~3 versus the amplitude~$q$ of the cosine potential energy.  (b)~The ratio $\varepsilon^{\rm gap}(q)/\varepsilon^{\rm gap}(q\to 0)$ versus~$q$.  The values of $\varepsilon^{\rm gap}(q\to0)/q^n$ are 1, 1/2, and 1/32 for $n = 1$, 2, and 3, respectively.}
\label{Fig:band_gaps_Mathieu}
\end{figure}

Figure~\ref{Fig:band_gaps_Mathieu}(a) shows the band gaps $\varepsilon^{\rm gap}/\pi^2$ versus the amplitude $q$ of the cosine potential at the first three Brillouin zone boundaries $ka/\pi = n =1$, 2, and~3. At the first Brillouin zone boundary with $n=1$ the band gap is proportional to~$q$ for small~$q$, but also contains higher-order $q$ contributions as further discussed in the following section.  Figure~\ref{Fig:band_gaps_Mathieu}(b) shows that the gaps for small $q$ are proportional to $q^n$ with numerical values $\varepsilon^{\rm gap}(q\to0)/(\pi^2q^n)  = 2$, 1/2, and 1/32 for $n=1,$ 2, and~3, respectively.  These values lead to the reduced energy gaps to lowest order in $u_1$ in Eq.~(\ref{Eqs:a1q}), respectively, given by
\bse
\bea
\varepsilon^{\rm gap}(q\to0, n=1) &=& u_1,\\
\varepsilon^{\rm gap}(q\to0, n=2) &=& \frac{1}{8\pi^2}u_1^2,\\
\varepsilon^{\rm gap}(q\to0, n=3) &=& \frac{1}{256\pi^4}u_1^3.
\eea
\ese
The value of $\varepsilon^{\rm gap}(q\to0)=u_1$ for $n=1$ agrees with previous calculations for a small-amplitude sinusoidal potential with an argument $G_1x$ where $G_1 = 2\pi/a$, called the nearly-free-electron model, see {\it e.g.}~\cite{Ashcroft1976, Hook2010, Kittel2008}.  The band gaps at Brillouin zone boundaries other than at $n = 1$ arise because a sinusoidal wave function containing the single reciprocal-lattice vector $G_1$ is not a solution of the Schr\"odinger equation~(\ref{Eq:Sch99}), which instead are Mathieu functions as discussed in the following section.

\begin{figure}
\includegraphics[width=3.in]{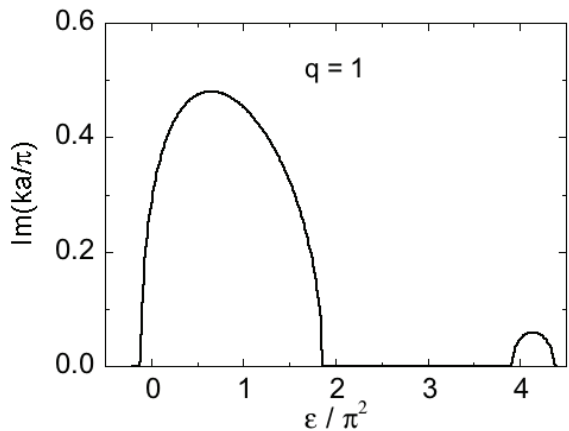}
\caption{Imaginary part Im$(ka/\pi)$ of the reduced electron wave vector $ka/\pi$ versus reduced energy $a_1 = \varepsilon/\pi^2$ for cosine potential-energy amplitude $q=1$.  For band states \mbox{Im$(ka/\pi) = 0$} whereas for energies in the band gaps \mbox{Im$(ka/\pi)>0$}.  The energy range plotted includes the band-gaps at the first and second Brillouin zone boundaries in Fig.~\ref{Fig:Matthieu_band_struct}(c).}
\label{Fig:Im_kaOnPi_VsEnergy_q1}
\end{figure}

\begin{figure}
\includegraphics[width=3.in]{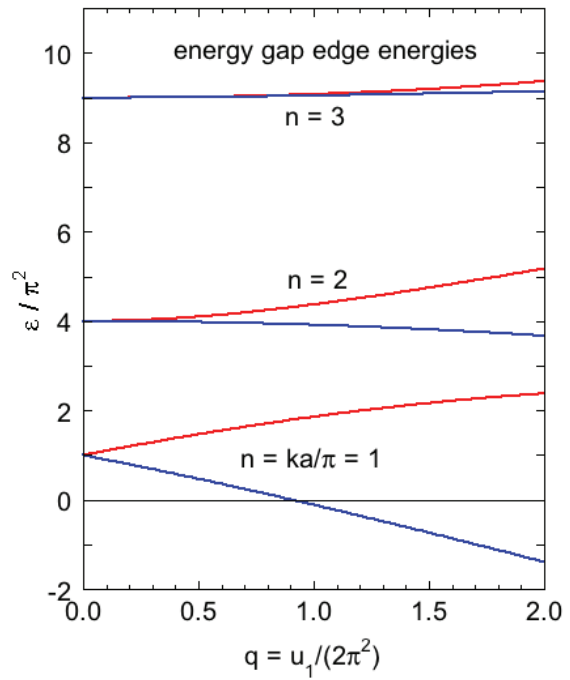}
\caption{Reduced energy $\varepsilon/\pi^2$ versus the amplitude~$q$ of the cosine potential energy at the upper (red) and lower (blue) edges of the energy gaps at the Brillouin zone boundaries with $n=ka/\pi = 1$, 2, and 3 as noted.}
\label{Fig:MathieuBandEdgesVSq}
\end{figure}

For energies in the band gaps, the electron wave vector $k$ in the extended-zone scheme is complex with the form $ka/\pi = n + {\rm Im}(ka/\pi)$, where Im denotes the imaginary part and $n$ is the Brillouin zone-boundary index associated with the energy gap.  Figure~\ref{Fig:Im_kaOnPi_VsEnergy_q1} shows Im$(ka/\pi)$ versus reduced energy $a_1 = \varepsilon/\pi^2$ for $q=1$, which illustrates that Im$(ka/\pi)$ increases from zero at each edge of a band gap, reaches a maximum value within the gap, and then decreases to zero at the upper edge of a band gap.  According to Eq.~(\ref{Eq:DispRelnFERedVar}), the value of $\varepsilon/\pi^2$ for free electrons is 1 for $n=1$ and~4 for $n=2$.  Figure~\ref{Fig:Im_kaOnPi_VsEnergy_q1} thus shows that the free-electron dispersion relation does not generally pass through the centers of the band gaps, which is not obvious from Fig.~\ref{Fig:Matthieu_band_struct}.  The reduced energies of the gap edges for the first three energy gaps are plotted versus~$q$ in Fig.~\ref{Fig:MathieuBandEdgesVSq}, which indeed show that the center of the gap decreases with increasing $q$ for the gap at $n=1$ and increases for the gaps at $n=2$ and~$n=3$, where the first two results are consistent with the behavior in Fig.~\ref{Fig:Im_kaOnPi_VsEnergy_q1}.

\section{\label{Sec:WaveFcns} Wave Functions}

For band states, the crystal momentum $k$ is real.  However, for states in the energy gaps $k$ is complex as discussed above.  In this section, the wave functions of propagating electron-band states are presented and discussed.

Equation~(\ref{Eq:psiFE}) for the wave function of the free electron can be written
\bea
\psi = A\cos(kx) + i B\sin(kx),
\label{Eq:PsiFE59}
\eea
where $A$ and~$B$ are real coefficients and $k$ is related to the energy~$E$ by Eq.~(\ref{Eq:EFE}).  The corresponding solution with the cosine potential energy included in the Schr\"odinger equation~(\ref{Eq:Sch99}) contains even- and odd-parity Mathieu functions denoted here by MC$[a_r(q),q,\pi x_a]$ and MS$[a_r(q),q,\pi x_a]$, respectively, which are both real functions where the last letters C and S refer to the even cosine and odd sine functions that the Mathieu functions reduce to when the cosine potential energy amplitude $q=0$ in Eq.~(\ref{Eq:Sch99}), and
\bea
a_r(q) = \frac{\varepsilon}{\pi^2}
\eea
according to Eq.~(\ref{Eq:a1Def}).  The corresponding traditional forms of the Mathieu functions are ce$_m(z,q)$ and se$_m(z,q)$, where here $m = r$ and $z = \pi x_a$.

To construct a wave function for the present problem analogous in form to that for a free electron in Eq.~(\ref{Eq:PsiFE59}), and which reduces to that form when $q=0$, we write
\bea
\psi(x_a) &=& A\ {\rm MC}[a_r(q),q,\pi x_a] + i B\ {\rm MS}[a_r(q),q,\pi x_a]\nonumber\\
&=& A\Big({\rm MC}[a_r(q),q,\pi x_a] + i \frac{B}{A}\ {\rm MS}[a_r(q),q,\pi x_a]\Big).\nonumber\\
\label{Eq:psiMathieu}
\eea

From Bloch's theorem~(\ref{Eq:psi(z)}) for a propagating electron, the function $f(x_a)$ periodic with the lattice is obtained from $\psi(x_a)$ according to
\bea
f(x_a) = e^{-i(ka/\pi)\pi x_a}\psi(x_a),
\label{Eq:fxa}
\eea
where $k$ is again the crystal momentum of the electron. Periodicity of $f(x_a)$ with the lattice requires
\bea
f(x_a+1) = f(x_a),
\eea
which for $x_a=0$ gives
\bea
f(1) = f(0),
\label{Eq:psi10}
\eea
We find that this criterion also leads to~\cite{{psipError}}
\bea
\frac{df(x_a)}{dx_a}(1) = \frac{df(x_a)}{dx_a}(0),
\eea
as illustrated in plots of $f(x_a)$ in Fig.~\ref{Fig:f_q1_r.25to1} below.  Inserting the expression in Eq.~(\ref{Eq:psiMathieu}) into~(\ref{Eq:fxa}) and applying boundary condition~(\ref{Eq:psi10}) to the resulting expression gives
\bea
\frac{B}{A} = i\frac{e^{i\pi r}{\rm MC}(x_a=0)-{\rm MC}(x_a=1)}{e^{i\pi r}{\rm MS}(x_a=0)-{\rm MS}(x_a=1)},
\label{Eq:BonA}
\eea
where for brevity only the $x_a$ dependencies of MC and MS are shown.  Equation~(\ref{Eq:BonA}) is then substituted into the second equation in Eq.~(\ref{Eq:psiMathieu}) to obtain $\psi$.  The remaining coefficient~$A$ is the normalization factor which is discussed further in the following.  

\subsection{Overview of the Wave Functions}

\begin{figure}
\includegraphics[width=3.4in]{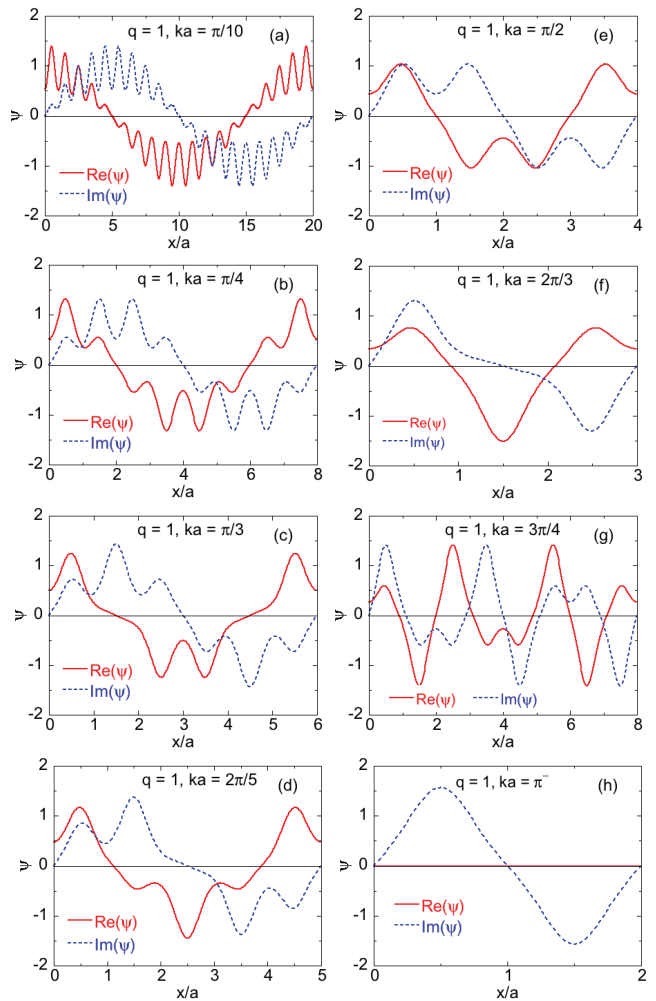}
\caption{Real and imaginary parts of the wave functions~$\psi$ versus position $x/a$ for amplitude~$q=1$ of the cosine potential energy for electron wave vectors $ka$ as listed in panels (a)--(h). The normalization factor $A$ in Eq.~(\ref{Eq:psiMathieu}) is set to unity.}
\label{Fig:psi_q1_rx.xx}
\end{figure}

Figure~\ref{Fig:psi_q1_rx.xx} shows an overview of the real and imaginary parts of the wave functions~$\psi$ versus $x/a$ for $q=1$ and eight values of the electron wave vector $ka$.  As discussed below, the electron probability density has an integrated value of unity in each unit cell of width~$a$ for each value of $ka$ for the coefficient $A=1$ in Eq.~(\ref{Eq:psiMathieu}).  The wavelengths of the wave functions in units of $a$ are $\lambda/a = m (2\pi/ka)$ where $m$ is the smallest positive integer for which $\lambda/a$ is an integer, which is the respective abscissa scale in Fig.~\ref{Fig:psi_q1_rx.xx} for each value of $ka$.  The figure shows dramatic deviations from the sinusoidal behavior that occurs when $q>0$.  However, for the smallest wave vector $ka = \pi/10$ in Fig.~\ref{Fig:psi_q1_rx.xx}(a), a modulation of the free-electron behavior in Eq.~(\ref{Eq:PsiFE59}) is seen due to the presence of the cosine potential.  This identification becomes less and less clear with increasing $ka$.

\begin{figure}
\includegraphics[width=3.4in]{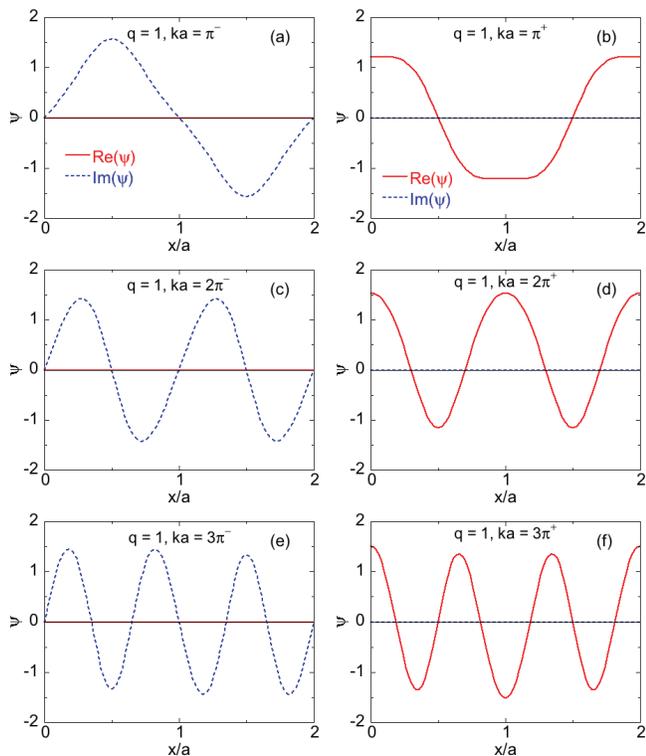}
\caption{Real and imaginary parts of the wave functions~$\psi$ versus position $x/a$ for amplitude~$q=1$ of the cosine potential energy for electron wave vectors at the Brillouin zone boundaries at the bottoms $ka^-$ (left panels) and tops $ka^+$ (right panels) of the energy gaps  as listed, corresponding to the tops and bottoms of the respective energy bands. The wave functions at the bottoms of the gaps are pure imaginary whereas at the tops of the gaps they are pure real, both corresponding to standing wave solutions.  The normalization factor $A$ in Eq.~(\ref{Eq:psiMathieu}) is unity.}
\label{Fig:psi_q1_r123}
\end{figure}

The wave function for $ka=\pi^-\ (ka=0.9999\pi)$ in Fig.~\ref{Fig:psi_q1_rx.xx}(h) is pure imagninary.  This wave vector is at the top of the lowest-energy band in the first Brillouin zone in Fig.~\ref{Fig:Matthieu_band_struct}(c).  This means that this wave function is a standing wave similar to the free-electron wave function in Eq.~(\ref{Eq:PsiFE59}) with $A=0$ resulting from Bragg reflection of the electron at the first Brillouin zone boundary.  Figures~\ref{Fig:psi_q1_r123}(a--f) show the real and imaginary parts of $\psi(x_a)$ at the bottoms and tops of the energy gaps at $ka=\pi,\ 2\pi$, and $3\pi$ for $q=1$. The wave functions at the tops of the gaps are all imaginary as in Fig.~\ref{Fig:psi_q1_rx.xx}(h) whereas the wave functions at the bottoms of the gaps are all real, reflecting the different natures of the respective standing waves.  In particular, considering the propagating waves traveling to the right in Eq.~(\ref{Eq:PsiFE59}), one could say that for waves at the tops of the energy gaps (bottoms of the energy bands), the Bragg-reflected waves traveling to the left interfere destructively with the incident waves resulting in wave-function nodes at the atomic positions, whereas the Bragg-reflected waves at the bottoms of the energy gaps interfere constructively with the incident waves resulting in antinodes at those positions.  A classical analogy is  transverse waves on a stretched string reflected from a fixed or free end which interfere destructively or constructively with the incident waves resulting in a node or antinode at the end, respectively.

\begin{figure}
\includegraphics[width=3.4in]{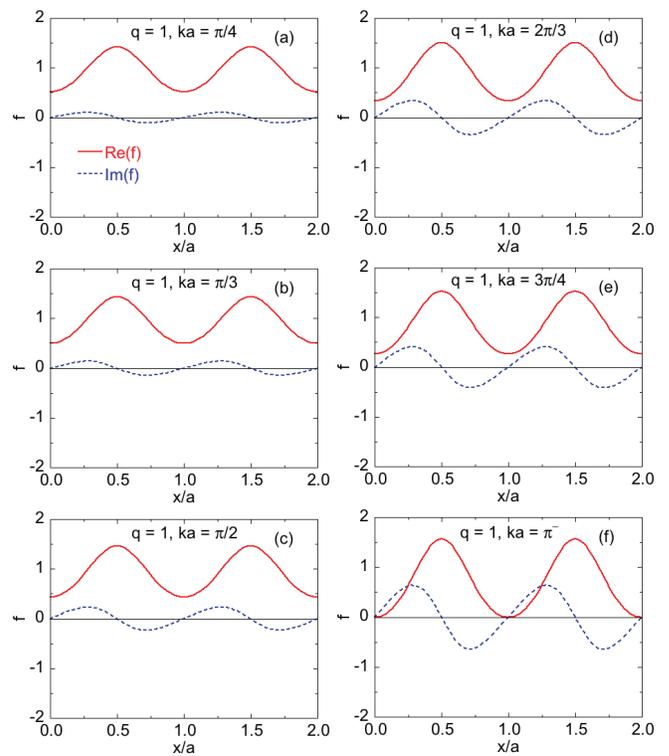}
\caption{Real (solid red curves) and imaginary (dashed blue curves) parts of the functions~$f(x_a)$ that are periodic with the lattice obtained from $\psi(x_a)$ using Eq.~(\ref{Eq:fxa}) for amplitude~$q=1$ of the cosine potential energy. The normalization factor $A$ in Eq.~(\ref{Eq:psiMathieu}) is set to unity, for which the $f(x_a)$ functions are normalized to unity over one unit cell.}
\label{Fig:f_q1_r.25to1}
\end{figure}

The real and imaginary parts of the periodic function $f(x_a)$ obtained from $\psi(x_a)$ using Eq.~(\ref{Eq:fxa}) are plotted in Fig.~\ref{Fig:f_q1_r.25to1} for six values of $ka/\pi$.  These plots are in the range $0\leq x_a \leq 2$, even though the period is unity, in order to illustrate the periodicity.  Note that for $ka=\pi^-$, $f(x_a)$ in Fig.~\ref{Fig:f_q1_r.25to1}(f) has both real and imaginary parts, whereas $\psi(x_a)$ for $ka=\pi^-$ in Fig.~\ref{Fig:psi_q1_rx.xx}(h) is pure imaginary.

\begin{figure}
\includegraphics[width=3.4in]{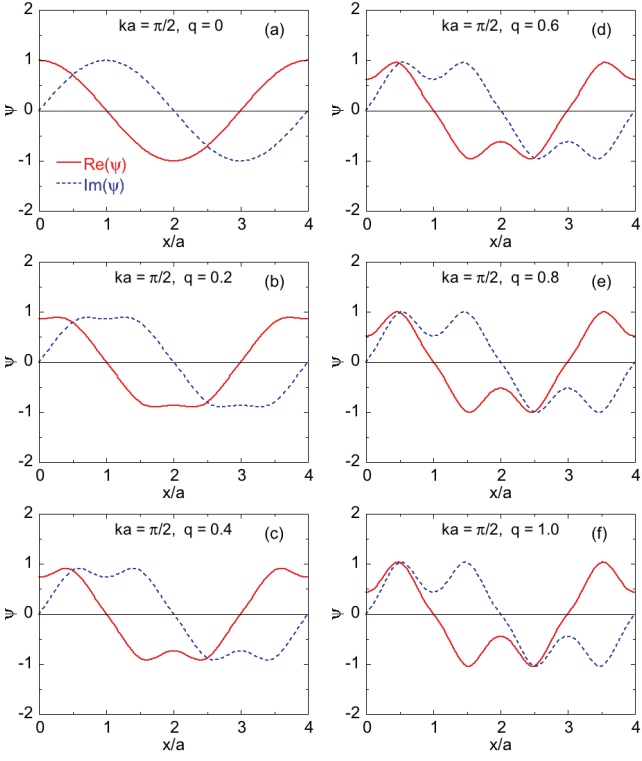}
\caption{Real and imaginary parts of the wave functions $\psi$ for $x/a=0$ to~1 and $ka = \pi/2$ with increasing $q$ from 0 to 1 in (a)--(f), respectively. The wavelength of $\psi$ in each case is $\lambda/a = 2\pi/ka = 4$ in the reduced-zone scheme.}
\label{Fig:Psi_r0.5_q0.24681}
\end{figure}

\begin{figure}
\includegraphics[width=2.5in]{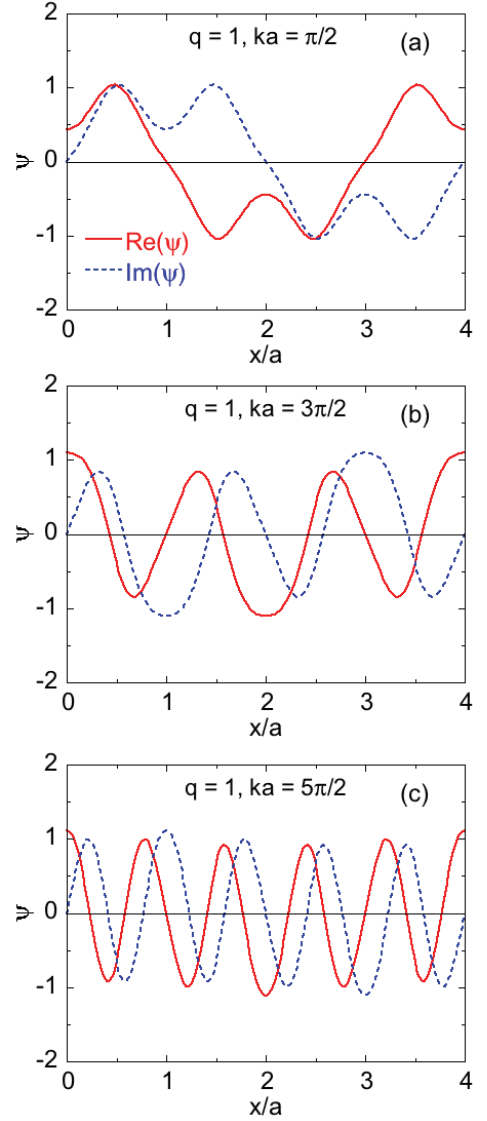}
\caption{Wave functions $\psi(x_a)$ for $q=1$ with (a)~$ka/\pi = 1/2$, (b)~$ka/\pi = 3/2$, and~(c)~~$ka/\pi = 5/2$ for the first three Brillouin zones in the extended-zone scheme, respectively.  The wavelength of $\psi$ in each case is $\lambda/a = 2\pi/ka = 4$ in the reduced-zone scheme.}
\label{Fig:Psi_q1_r0.5_1.5_2.5}
\end{figure}

Figure~\ref{Fig:Psi_r0.5_q0.24681} shows the influence of the amplitude $q$ of the cosine potential on the wave functions for a wave vector $ka=\pi/2$ in the middle of the first Brillouin zone for the range $q = 0$ to $q = 1$.  Even a relatively small value $q=0.2$ leads to a significant distortion of the wave function compared to sinusoidal wave function of the free electron in Fig.~\ref{Fig:Psi_r0.5_q0.24681}(a) for which $q=0$.  Figures~\ref{Fig:Psi_q1_r0.5_1.5_2.5}(a--c) show $\psi(x_a)$ for energy bands 1, 2, and 3 at wave vectors $ka=\pi/2,\ 3\pi/2$, and $5\pi/2$ in the extended-zone scheme, respectively, which are equivalent wave vectors in the reduced-zone scheme.  The wave functions are seen to become more free-electron-like with increasing energy.


\subsection{Fourier Components of the Wave Functions}

\begin{figure}
\includegraphics[width=3.4in]{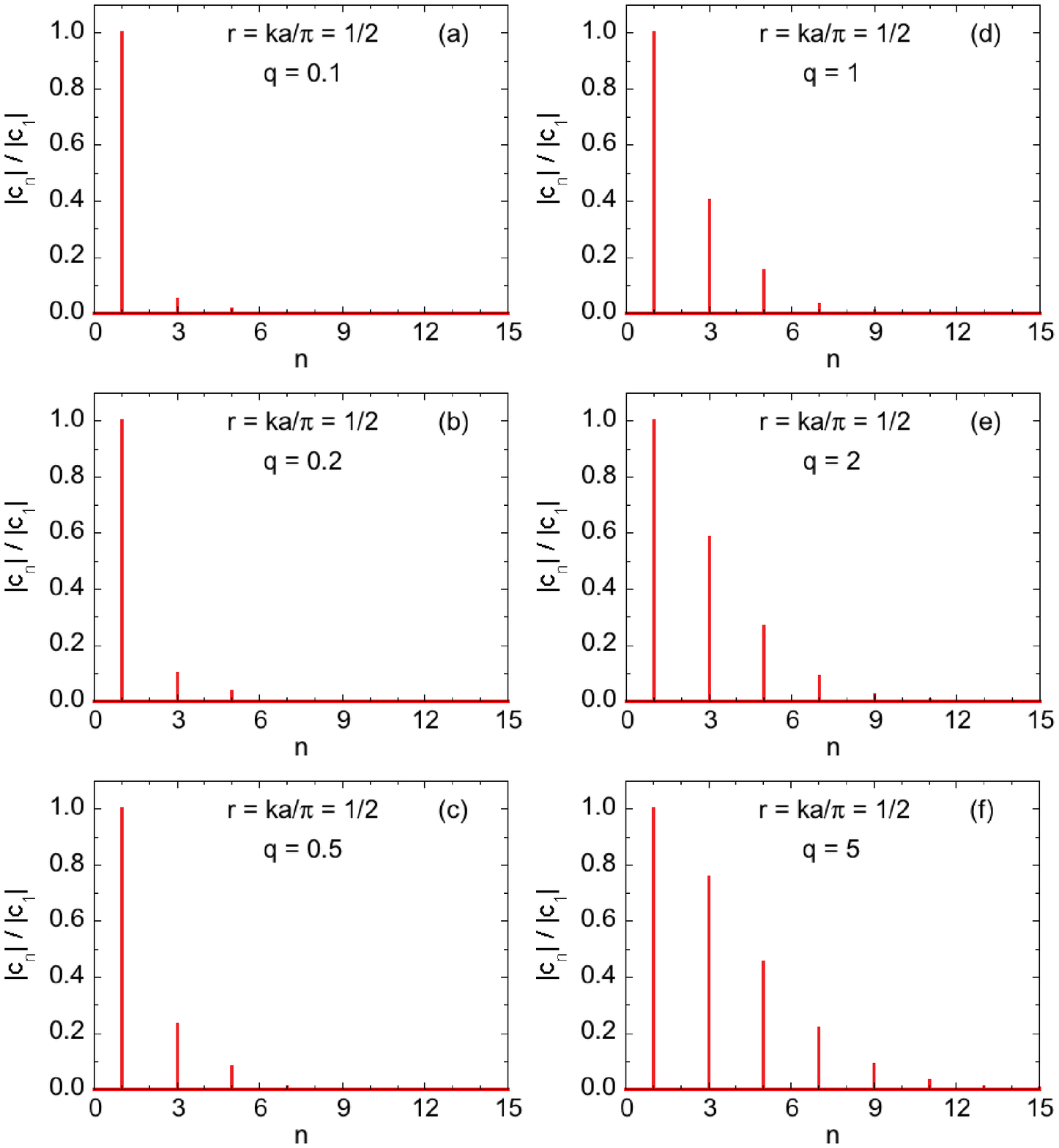}
\caption{Ratio $|c_n|/|c_1|$ of the magnitudes of the Fourier series coefficients in Eq.~(\ref{Eq:psi(x)Gnx}) for reduced electron wave vector \mbox{$r = ka/\pi = 1/2$} versus the order $n$ of the reciprocal-lattice vector $G_n = n2\pi/a$  for the $q$ values listed.  For this value of $ka/\pi$ only odd harmonics of $G_1$ occur for each value of $q$ and they each become stronger as $q$ increases. The same Fourier-coefficient ratios occur for negative~$n$ which can be obtained by reflecting the spectra about $n=0$.}
\label{Fig:cnOnc1qX.Xr0.5}
\end{figure}

\begin{figure}
\includegraphics[width=3.4in]{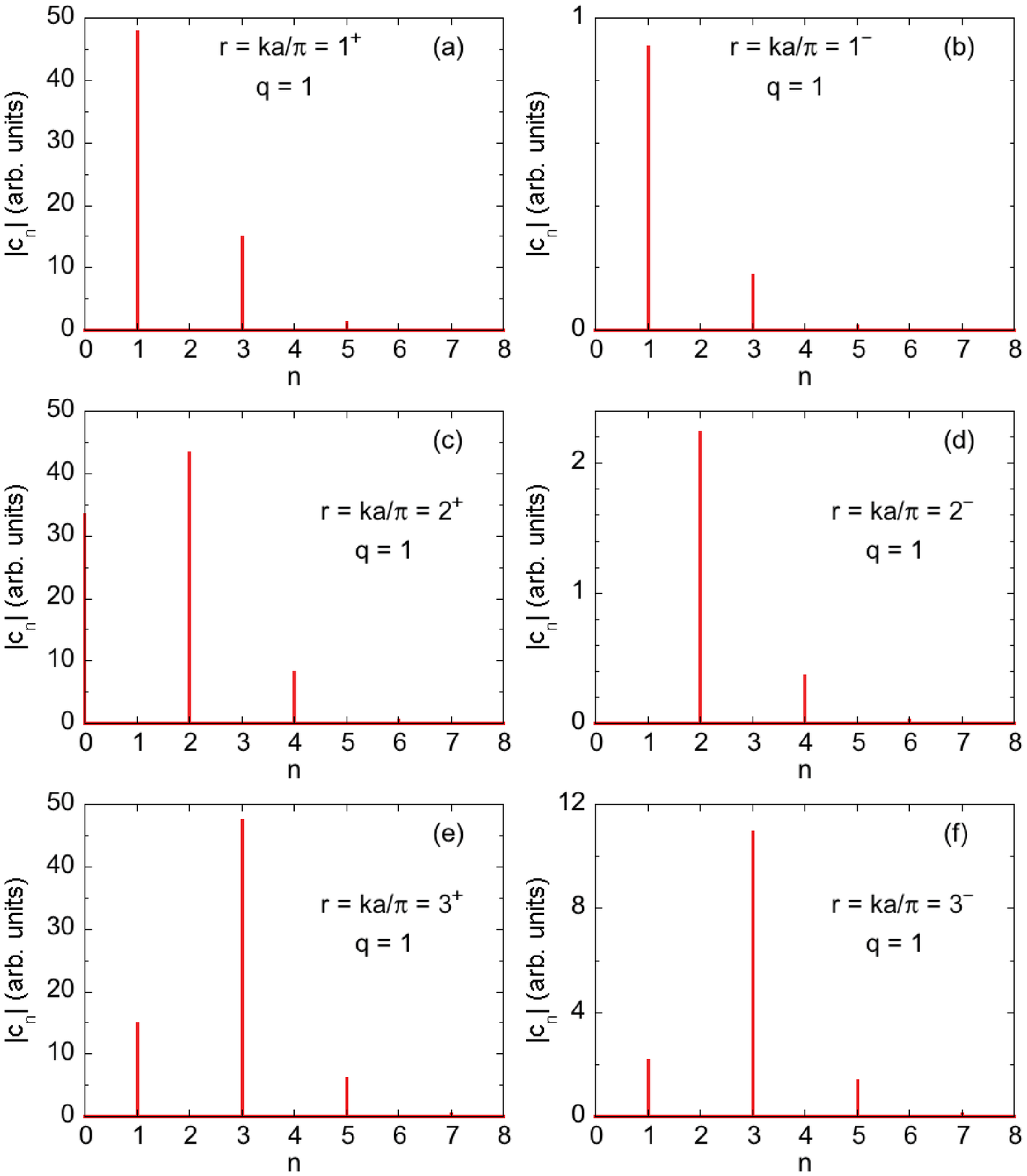}
\caption{Unnormalized amplitudes $|c_n|$ versus $n = G_na/2\pi$ of the Fourier spectra of the respective nonzero real or imaginary parts of the wave functions  in Fig.~\ref{Fig:psi_q1_r123} near the tops [(a), (c), (e)] and bottoms [(b), (d), (f)] of the first three energy gaps at $n = 1$, 2, and 3, respectively.  Note that the relative amplitudes of the peaks for a given $r^+$ are close those of the respective amplitudes of the peaks for $r^-$, that the scales of the ordinates in each of (b), (d), and~(f) are different, these scales are much smaller than in (a), (c), and~(e), and a zero-frequency (dc) component occurs in~(c).  The same expansion coefficients occur for negative~$n$ which can be obtained by reflecting the spectra about $n=0$.}
\label{Fig:cn_q1_rNplusminus}
\end{figure}

In the limit $q\to0$ for which the Schr\"odinger equation gives free-electron wave functions, the Mathieu functions are respectively just sine and cosine waves with argument $kx$ and energy $E_k = \hbar^2k^2/2m$.  However, with increasing $q$, reciprocal-lattice vectors
\bea
G_n = n2\pi/a
\eea
appear in the wave function.   A wave function of an electron in a periodic lattice in one dimension can be expressed as a Fourier-series expansion in terms of $G_n$ given by
\bea
\psi(x) = \sum_{n=1}^\infty c_n e^{i G_n x},
\label{Eq:psi(x)Gnx}
\eea
where $c_n$ is a complex coefficient.  One can therefore determine the amplitudes $c_n$ associated with a particular wave function with a specific value of $r = ka/\pi$ via Fourier analysis.  Below the Fourier amplitude spectrum of the real part of $\psi(x)$ for $r = ka/\pi = 1/2$ is calculated (the spectrum of the imaginary part is the same except for values of $r$ corresponding to a Brillouin zone boundary with $r=1,2,\ldots$).

Due to the discrete nature of the allowed electron wave vectors~$k$ in a ring containing a finite number $N$ of ions under periodic boundary conditions, the allowed values of $ka/\pi$ must be rational numbers.  We find that if an irrational value such as $ka/\pi=1/\pi$ is used, a continuous Fourier spectrum is obtained instead of the discrete spectrum required by Eq.~(\ref{Eq:psi(x)Gnx}). We only plot the spectra for positive values of Brillouin zone-boundary indices~$n = G_na/2\pi = ka/\pi$, since the spectra for negative~$n$ are mirror images of the positive-$n$ values.  Figure~\ref{Fig:cnOnc1qX.Xr0.5} shows the ratios $|c_n|/|c_1|$ versus~$n$ for $r = ka/\pi = 1/2$ and $q = 0.1$ to~5.  As $q$ increases, the width of the visible Fourier spectrum increases.  However, we show below that many higher-order Fourier components are also present but with amplitudes smaller than can be seen in Fig.~\ref{Fig:cnOnc1qX.Xr0.5}.

The Fourier spectra for the electron-band wave functions in Fig.~\ref{Fig:psi_q1_r123} at energies close to the tops and bottoms of the first three energy gaps are shown in Fig.~\ref{Fig:cn_q1_rNplusminus}, where the unnormalized spectra were respectively calculated from either the real or imaginary part of $\psi$ depending on which was nonzero.  The spectra versus reciprocal-lattice index $n = r = ka/\pi = G_na/2\pi$ show interesting features.  First, the strongest component is at $ka/\pi = n$ which specifies which gap is considered in the extended-zone scheme.  Second, there exist Fourier components with other reciprocal-lattice wave vectors with smaller amplitude than the one at $n=r$.  Third, for $r = 2^+$, a Fourier component occurs at $n = 0$ corresponding to a constant vertical shift in the wave function, in agreement with the upward shift of the average of the wave function in Fig.~\ref{Fig:psi_q1_r123}(d).  Fourth, the Fourier components of the wave functions at energies slightly less than the bottoms of the gaps  in Figs.~\ref{Fig:cn_q1_rNplusminus}(b,d,f) are much smaller than those at energies slightly greater than the tops of the gaps in Figs.~\ref{Fig:cn_q1_rNplusminus}(a,c,e) as is apparent from the corresponding wave functions in Fig.~\ref{Fig:psi_q1_r123}, although the relative amplitudes of the respective peaks are similar.  Finally, one might have expected that the only peak in the Fourier spectra for wave vectors infinitesimally close to $n = ka/\pi$ would be at $G_n$, but the data in Fig.~\ref{Fig:cn_q1_rNplusminus} show that this is not the case.

\begin{figure}
\includegraphics[width=3.in]{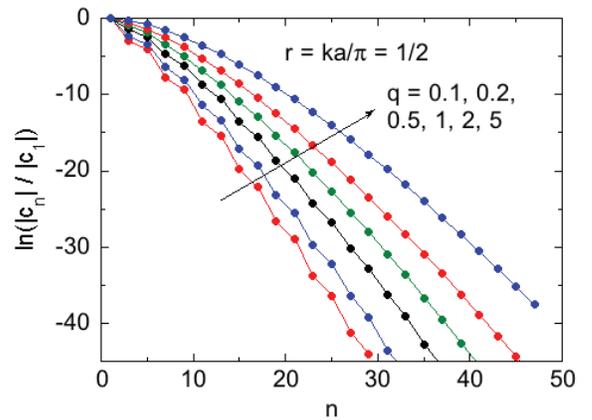}
\caption{Natural logarithm of the ratio $|c_n|/|c_1|$ of the expansion coefficients in Eq.~(\ref{Eq:psi(x)Gnx}) for reduced electron wave vector $r = ka/\pi = 1/2$ versus the order $n$ of the reciprocal-lattice vector $G_n = n2\pi/a$  for the $q$ values listed.  For $ka/\pi=1/2$ only odd harmonics of $G_n/2$ occur for each value of $q$.  The lines are guides to the eye.}
\label{Fig:LogcnOnc1qX.Xr0.5}
\end{figure}

\begin{figure}
\includegraphics[width=3.3in]{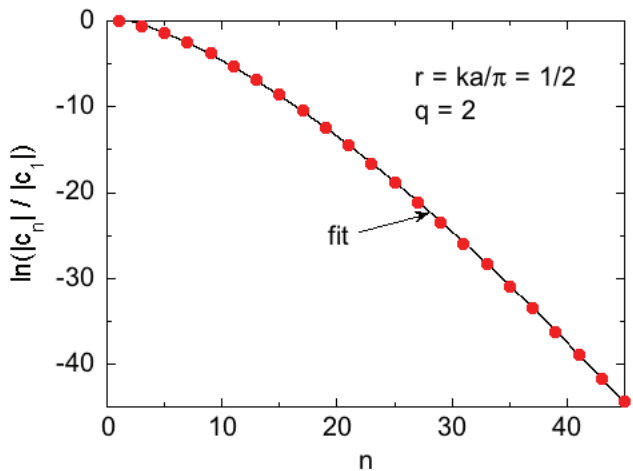}
\caption{Natural logarithm of the ratio $|c_n|/|c_1|$ of the expansion coefficients in Eq.~(\ref{Eq:psi(x)Gnx}) versus the order $n$ of the reciprocal lattice vector $G_na/2 = n\pi$ (solid red circles) for amplitude $q=2$ of the cosine potential with $ka/\pi = 1/2$.  The fit by Eqs.~(\ref{Eqs:Fit2}) is shown as the black curve.}
\label{Fig:LogcnOnc1q2r0.5Fit2}
\end{figure}

Figure~\ref{Fig:LogcnOnc1qX.Xr0.5} illustrates the $q$ dependence of $\ln(|c_n|/|c_1|)$ versus~$n$ for $ka/\pi = 1/2$ with $q=0.1$ to~5.  The plots show that the dependence on~$n$ weakens as $q$ increases, consistent with Fig.~\ref{Fig:cnOnc1qX.Xr0.5}.  We also see from Fig.~\ref{Fig:LogcnOnc1qX.Xr0.5} that (i)~only the odd harmonics of the fundamental reciprocal-lattice wave vector $G_1$  occur for $ka/\pi = 1/2$; (ii)~the dependence of $\ln(|c_n|/|c_1|)$ on~$n$ is clearly seen to be sawtooth-shaped for the smaller $q$ values; and (iii)~$|c_n|/|c_1|$ falls off faster than $e^{-B n}$ versus~$n$, where $B$ is a positive constant.

Figure~\ref{Fig:LogcnOnc1q2r0.5Fit2} shows $\ln(|c_n|/|c_1|)$ versus~$n$ from $n=1$ to $n=45$ for $ka/\pi = 1/2$ with fixed cosine potential amplitude $q=2$ [see Fig.~\ref{Fig:cnOnc1qX.Xr0.5}(e)]. An excellent fit of the data shown by the black curve in Fig.~\ref{Fig:LogcnOnc1q2r0.5Fit2} was obtained by the expression
\bse
\label{Eqs:Fit2}
\bea
\ln\left(\frac{|c_n|}{|c_0|}\right) = A + B n^c,
\eea
where
\bea
A = 0.52(8),\quad B = -0.184(5), \quad c = 1.445(7),
\eea
\ese
with a goodness-of-fit parameter $R^2 = 0.99992$.  

\section{\label{PJ} Probability Density and Probability Current}

\begin{figure*}
\includegraphics[width=7in]{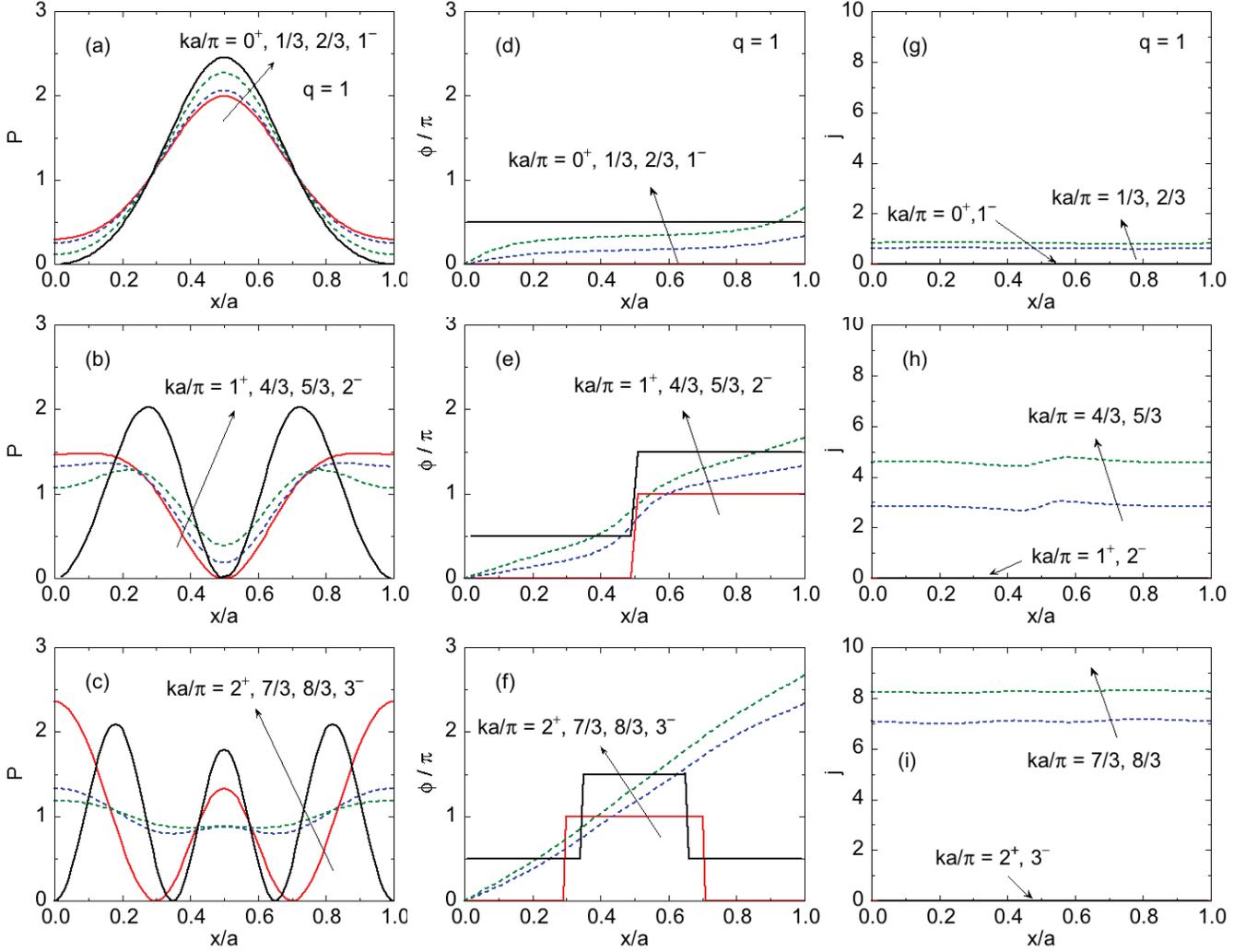}
\caption{(a)--(c) Probability density $P$ in Eq.~(\ref{Eq:Pxa}), (d)--(f) phase $\phi$ of the wave function in Eq,~(\ref{Eq:psiCe}), and (g)--(i) the reduced probability current $j$ in Eq.~(\ref{Eq:j(xa)Def}) verus position $x_a$ for four wave vectors in each of the first three Brillouin zones, respectively.}
\label{Fig:P_q1}
\end{figure*}

The probability density
\bea
P(x_a) = \psi^\ast(x_a)\psi(x_a) = f^\ast(x_a)f(x_a)
\eea
is the (real) probability per unit length along the $x$~axis. $P(x_a)$ is the same for $\psi(x_a)$ and $f(x_a)$ as is apparent from Eq.~(\ref{Eq:fxa}).  Figures~\ref{Fig:P_q1}(a)--(c) show plots of $P$ versus $x_a\equiv x/a$ over one unit cell for cosine potential-energy ampliutude $q=1$ obtained from wave functions such as  discussed in the previous section for representative values of $ka$ in the first, second, and third Brillouin zones.  For each $ka$, $P(x_a)$ is seen to be periodic versus $x_a$ with a period of unity as required.  With the choice $A=1$ for the normalization factor, $P(x_a)$ integrated over one unit cell is unity for each value of $ka$.  For each Brillouin zone, nodes in $P$ are seen at $x_a=0$ and~1 at the top of the electron band and an antinode at the bottom, consistent with the nodes and antinodes in the respective wave functions at $x_a=0$ and~1 in Fig.~\ref{Fig:psi_q1_r123}.

The probability current $J(x)$ in one dimension is the rate at which probability flows past position~$x$ and has dimensions of 1/time.  It is obtained from the time-dependent Schr\"odinger equation and is given in one dimension by the real function~\cite{Griffiths2015}
\bea
J(x) = \frac{i\hbar}{2m}\left(\psi \frac{d\psi^\ast}{d x} -\psi^\ast \frac{d\psi}{d x} \right).
\label{Eq:J1}
\eea
If we write
\bea
\psi(x_a) = C(x_a) e^{i\phi(x_a)},
\label{Eq:psiCe}
\eea
where $C(x_a)$ is real with dimensions of $\sqrt{1/{\rm length}}$, then Eq.~(\ref{Eq:J1})  becomes
\bea
J(x_a) = \frac{\hbar}{ma}{\rm Im}\left(\psi^\ast\frac{d\psi}{d x_a}\right),
\label{Eq:JImPsi}
\eea
where again $x_a\equiv x/a$.  The probability density is
\bea
P(x_a) = C^2(x_a),
\label{Eq:Pxa}
\eea
so the probability current~(\ref{Eq:JImPsi}) becomes
\bea
J(x_a) =\frac{\hbar}{ma}P(x_a)\frac{d\phi(x_a)}{dx_a},
\label{Eq:J2}
\eea
where $\phi(x_a)$ is the phase of the wave function in Eq.~(\ref{Eq:psiCe}).  For plotting purposes a reduced probability current is defined as
\bea
j(x_a) =  \frac{ma}{\hbar}J(x_a) = P(x_a)\frac{d\phi(x_a)}{dx_a}.
\label{Eq:j(xa)Def}
\eea

Figures~\ref{Fig:P_q1}(d--f) show plots of $\phi/\pi$ versus~$x_a$ for the same sets of parameters in Figs.~\ref{Fig:P_q1}(a--c), respectively. The reduced probability current $j$ versus $x_a$ obtained by multiplying $P(x_a)$ by the slope of $\phi(x_a)$ according to Eq.~(\ref{Eq:j(xa)Def}) is plotted in Figs.~\ref{Fig:P_q1}(g--i) for the respective variables in Figs.~\ref{Fig:P_q1}(a--c).  One immediately sees from Figs.~\ref{Fig:P_q1}(g--i) that $j(x_a)=0$ for states at the tops and bottoms of the energy bands because for these states $d\phi(x_a)/dx_a = 0$ from Figs~\ref{Fig:P_q1}(d--f).  This is consistent with our observation from Fig.~\ref{Fig:psi_q1_r123} that for such crystal momenta the wave functions are standing waves, which consist of waves with equal amplitudes but with crystal momenta in opposite directions, respectively, resulting from Bragg reflection of the electron waves from the ions producing the sinusoidal potential energy seen by the electron.  In addition, it is clear that the probability current increases with increasing $ka$ in the extended-zone sheme, which occurs because the electron-band energies correspondingly increase.

\section{\label{Sec:CentEq} Band Structure from the Central Equation}

In one dimension, the potential energy in Eq.~(\ref{Eq:U(x)1}) can be written
\bea
U(x) = \sum_G U_{G} e^{i G x}.
\label{Eq:Usum}
\eea
In the present paper, $U(x)$ in Eq.~(\ref{Eq:U(x)}) is real with a single Fourier component
\bea
G_1 &=& \pm 2\pi/a, \label{Eqs:GnU}\\
U_{G} &=&  U_1/2\equiv U.\nonumber 
\eea
The wave function for a given real crystal momentum $k_x >0 \equiv k$ can be written as the Fourier series
\bea
\psi_k(x) &=& \sum_{G} c_k e^{i(k-G)x},\label{Eq:psiGka}\\
G &=& \pm n\frac{2\pi}{a},\nonumber
\eea
where $\psi_k(x)=\psi_{k+G}(x)$ and $k$ is in the first Brillouin zone\@.  Examples of the magnitudes $|c_n|/|c_1|$ and $|c_n|$ of the Fourier components are plotted above versus~$n$ in Figs.~\ref{Fig:cnOnc1qX.Xr0.5} and~\ref{Fig:cn_q1_rNplusminus}, respectively.

Substituting Eqs.~(\ref{Eq:Usum}) and~(\ref{Eq:psiGka}) into the Schr\"odinger equation gives the so-called central equation~\cite{Kittel2008}
\bea
(E_k-E)c_k + \sum_G U_G c_{k-G} &=& 0,\label{Eq:central}\\
&&\hspace{-1.5in}E_k = \frac{\hbar^2k^2}{2m},\nonumber
\eea
which is the discrete Fourier transform of the Schr\"odinger equation in crystal-momentum space, where $E_k$ is the free-electron dispersion relation, and $k$ is again restricted to the first Brillouin zone.  For our 1D case, $U_G=U$ and $G = 2\pi/a$ with both positive and negative values $\pm G$, so Eq.~(\ref{Eq:central}) becomes
\bea
U c_{k+G} + (E_k-E)c_k +  U c_{k-G} = 0.
\label{Eq:UEU}
\eea
Since any multiple of $\pm G$ can be added to a particular $k$ with the same energy (repeated-zone scheme), there are in theory an infinite number of such equations.  However, Figs.~\ref{Fig:cnOnc1qX.Xr0.5} and~\ref{Fig:cn_q1_rNplusminus} demonstrate that only a small number of $G$ values are generally present with significant amplitudes.  As the amplitude $q$ of the cosine potential increases, so does the number of multiples of $G$ necessary to reproduce the wave functions such as in Fig.~\ref{Fig:Psi_r0.5_q0.24681}.  Here we are interested in seeing how the dispersion relations obtained from the central equation~(\ref{Eq:central}) depend on the number of included $G$ values for comparison with the numerically-exact solutions in Fig.~\ref{Fig:Matthieu_band_struct}.  According to Eq.~(\ref{Eq:UEU}), each value of $c_k$ is only coupled to two other values $c_{k+G}$ and $c_{k-G}$, etc., which simplifies calculation of the dispersion relations.

The five lowest-order equations in $G$ derived from Eq.~(\ref{Eq:UEU}) can be written
\begin{widetext}
\begin{gather}
\begin{pmatrix}
&&&\vdots&&\\
U 		& E_{k+2G} - E 	& 	U 			& 	0 				& 	0 				& 0 		& 0	\\
0		&	U 			& 	E_{k+G} - E 	& 	U 				& 	0 				& 0 		& 0 	\\
0		&	0			&	U 			& 	E_{k} - E 		& 	U 				& 0 		& 0  \\
0		& 	0			&	0			&	U 				& 	E_{k-G} - E 		& U 		& 0  \\
0		&	0			&	0			&	0				&	U 				& 	E_{k-2G} - E 		& U\\ 
&&&\vdots&&\\
\end{pmatrix}
\begin{pmatrix}
\vdots\\
c_{k+3G}\\
c_{k+2G}\\
c_{k+G}\\
c_{k}\\
c_{k-G}\\
c_{k-2G}\\
c_{k-3G}\\
\vdots\\
\end{pmatrix}
=0,
\end{gather}
where~$G\equiv |G_1| = 2\pi/a$ from Eq.~(\ref{Eqs:GnU}).

In order to solve for the energies~$E$, this $7\times 5$ matrix is reduced to a $5\times 5$ square matrix by eliminating the first and last columns of the matrix and the top and bottom entries of the column vector, yielding
\begin{gather}
\begin{pmatrix}
E_{k+2G} - E 	& 	U 			& 	0 				& 	0 				& 0 \\
U 			& 	E_{k+G} - E 	& 	U 				& 	0 				& 0 \\
0			&	U 			& 	E_{k} - E 		& 	U 				& 0 \\
0			&	0			&	U 				& 	E_{k-G} - E 		& U  \\
0			&	0			&	0				&	U 				& E_{k-2G} - E \\ 
\end{pmatrix}
\begin{pmatrix}
c_{k+2G}\\
c_{k+G}\\
c_{k}\\
c_{k-G}\\
c_{k-2G}\\
\end{pmatrix}
=0,
\label{Eq:Gmax2Mat}
\end{gather}
\end{widetext}

where, {\it e.g.},
\bea
E_{k+G} = \frac{\hbar(k+G)^2}{2m}.
\eea

\begin{figure}
\includegraphics[width=3.3in]{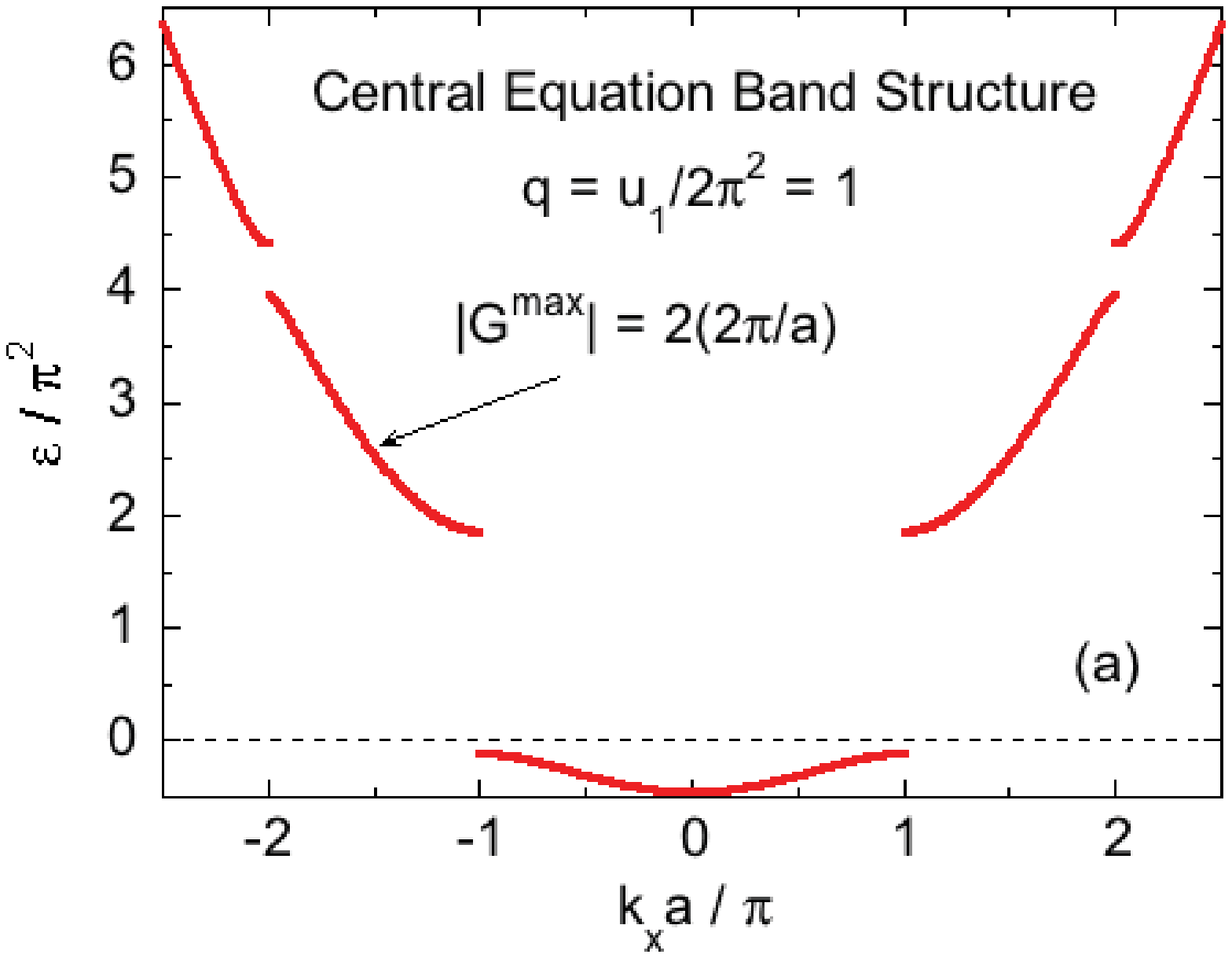}
\includegraphics[width=3.3in]{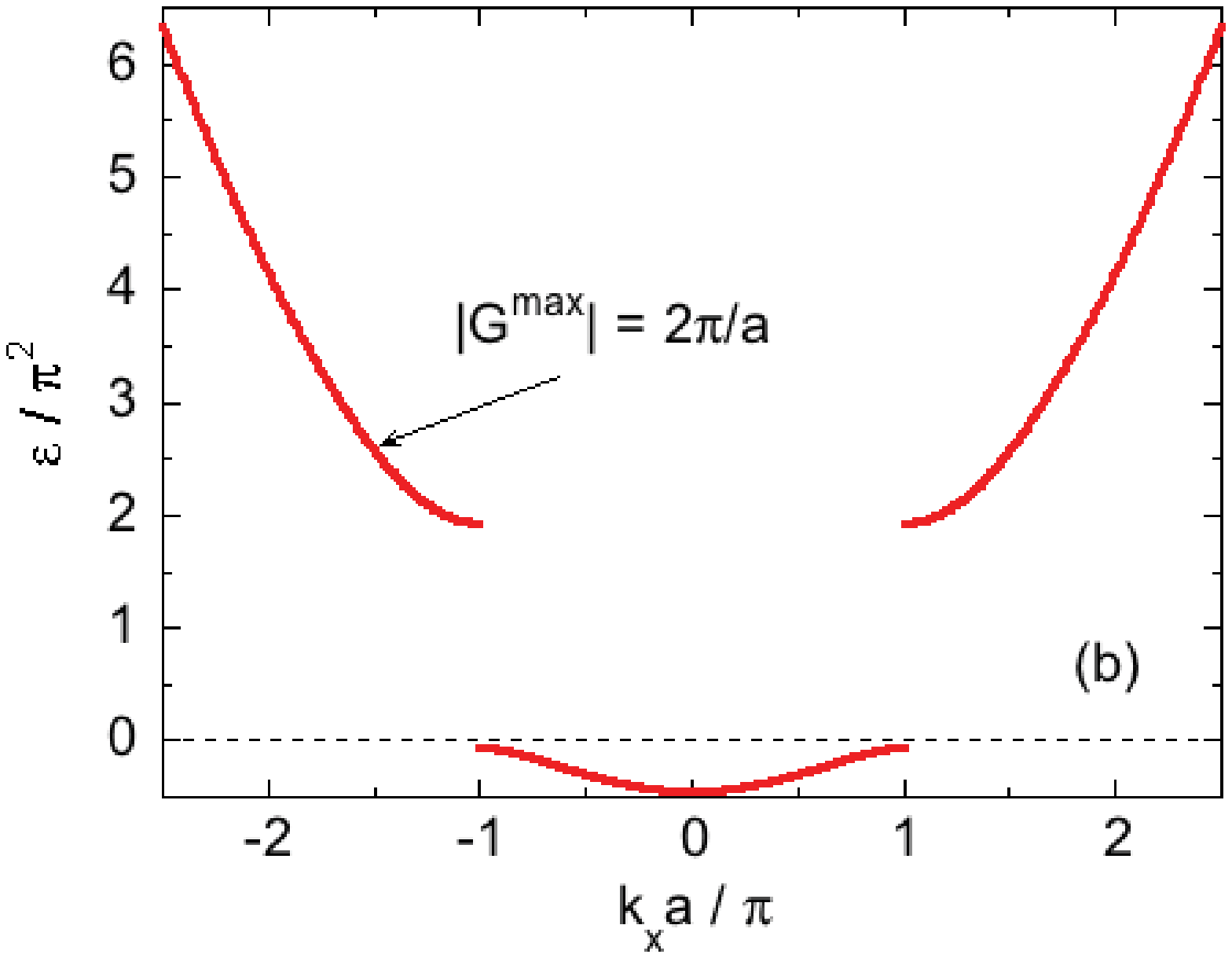}
\caption{Band structure obtained using the central equation~(\ref{Eq:central}) and matrix equation~(\ref{Eq:Gmax2Mat}) for cosine potential amplitude $q=1$ and maximum reciprocal-lattice vectors (a)~$G^{\rm max} = 2(2\pi/a)$ and (b)~$G^{\rm max} = 2\pi/a$. Reducing $G^{\rm max}$ from $4\pi/a$ to $2\pi/a$ is seen to eliminate the band gaps at $k_xa/\pi = \pm 2$.}
\label{Fig:CentEqnBSq1Gmax}
\end{figure}

In order to obtain nonzero solutions for the $c_{k\pm nG}$ coefficients, the determinant of the matrix must vanish, which yields the band structure $E(k)$.  Figure~\ref{Fig:CentEqnBSq1Gmax}(a) shows the band structure in the extended-zone scheme obtained from Eq.~(\ref{Eq:Gmax2Mat}) for $q=1$, where we have normalized the axes to agree with those in Fig.~\ref{Fig:Matthieu_band_struct}(c).  We find that the lowest-energy band gap at $k_xa/\pi=\pm1$ agrees to five significant figures with that obtained in Fig.~\ref{Fig:Matthieu_band_struct}(c) from numerically-exact calculations, whereas the second gap at $k_xa/\pi=\pm2$ is about 1\% too large.  When the matrix in Eq.~(\ref{Eq:Gmax2Mat}) is reduced to $3\times3$ to only take into account the reciprocal lattice vectors $G = \pm2\pi/a$, the energy gaps at $k_xa/\pi=\pm2$ disappear in the derived $E(k)$ relation as shown in Fig.~\ref{Fig:CentEqnBSq1Gmax}(b), and the energy gaps at $k_xa/\pi=\pm1$ are too large by about 1\% compared to the gaps in Fig.~\ref{Fig:Matthieu_band_struct}(c).  We conclude that the overall agreement of the band structure obtained for $q=1$ within the energy range in Fig.~\ref{Fig:CentEqnBSq1Gmax}(a) from the central equation, as obtained above, with the numerically-exact band structure in Fig.~\ref{Fig:Matthieu_band_struct}(c) is quite good for the energy range in Fig.~\ref{Fig:CentEqnBSq1Gmax}.


\section{\label{Sec:Conclusion} Concluding Remarks}

The band structure of noninteracting electrons in a one-dimensional metallic solid with a sinusoidal potential containing the first reciprocal-lattice vector has been discussed in the past in the nearly-free-electron approximation, where only the lowest-order contribution of the potential is discussed. The availability of numerically-exact solutions to the Mathieu Schr\"odinger equation has allowed far more information to be obtained about the band structure and wave functions.

The new results presented here include the dependence of the band structure on the amplitude of the sinusoidal potential, detailed wave functions and probability densities versus position and crystal-momentum~$k$ with a discussion of the normalization factor, Fourier-series analyses of the wave functions to show the amplitudes of the reciprocal-lattice components $n 2\pi/a$ and analyses of these components from $n=1$ to $n=45$, the probability currents associated with the electron bands in the first, second, and third Brillouin zones, and a comparison of the band structure with that obtained from the central equation.  An important result is that the occurrence of energy gaps in the band structure at wave vectors $k=G_n/2$ with $n>1$  in the extended-zone scheme does not require the presence of those wave vector components in the potential energy.

The sinusoidal potential is more realistic than the Kronig-Penney Dirac-comb model for one-dimensional metals~\cite{Kronig1931} often used an an introduction to students of the band structure of solids.

\acknowledgments

This work was supported by the U.S.\ Department of Energy, Office of Basic Energy Sciences, Division of Materials Sciences and Engineering.  Ames Laboratory is operated for the U.S.\ Department of Energy by Iowa State University under Contract No.~DE-AC02-07CH11358.


\begin{thebibliography}{99}
\bibitem{psipError} In version 11 of {\tt Mathematica} used in this paper, there is a bug in the program when calculating the derivatives $\partial {\rm MC}(x_a)/\partial  x_a$ and $\partial  {\rm MS}(x_a)/\partial  x_a$ as given by the built-in functions MathieuCPrime[$a_r(q),\ q,\, \pi x_a$]  and MathieuSPrime[$a_r(q),\ q,\, \pi x_a$], respectively.  They are both a factor of 2 too small. This bug was taken into account when calculating $d\psi(x_a)/dx_a$.
\end{thebibliography}
\end{document}